\documentstyle[preprint,aps]{revtex}
\begin{document}
\tighten
\newcommand{\beq}{\begin{equation}}
\newcommand{\eeq}{\end{equation}}
\newcommand{\eq}[1]{eq.(\ref{#1})}
\draft
\preprint{PSU/TH/161}
\title {NEW CORRECTIONS OF ORDER $\alpha^3(Z\alpha)^4$ AND
$\alpha^2(Z\alpha)^6$ TO
THE LAMB SHIFT}
\author {Michael I. Eides \cite{emaile}}
\address{Petersburg Nuclear Physics Institute,\\
Gatchina, St.Petersburg 188350, Russia}
\author {Howard Grotch \cite{emailg}}
\address{Department of Physics, Pennsylvania State University,\\
University Park, Pennsylvania 16802, USA\medskip}
\date{June 2}

\maketitle

\begin{abstract}
Two corrections to the Lamb shift, induced by the polarization operator
insertions in the external photon lines are calculated.
\end{abstract}

\pacs{PACS numbers:  12.20.Ds, 31.30.Jv}

After completion of calculations of all corrections of order
$\alpha^2(Z\alpha)^5$ to the Lamb shift
\cite{ego,eg,eg4,eg5,eksl1,eksl2,es,pach1,pach2} only unknown corrections of
orders $\alpha^3(Z\alpha)^4$, $\alpha^2(Z\alpha)^6$ and $\alpha(Z\alpha)^7$
may be large enough to deserve any phenomenological consideration (see, e.g.,
discussion in \cite{es}). We present below a calculation of
contributions to the Lamb shift induced by three and two loop
polarization insertions in the external photon lines.

\section{Correction of Order $\alpha^3(Z\alpha)^4$ Induced by the Three-Loop
Vacuum Polarization}

In a recent remarkable work \cite{bb} P. A. Baikov and D. J. Broadhurst have
calculated analytically three leading terms in the low frequency asymptotic
expansion of the three-loop polarization operator. For the
numerical factor in the leading term $p_3(\alpha/\pi)^3(q^2/m^2)$ they
obtained

\beq
4p_3\equiv C_1=
-\frac{325805}{93312}+\frac{23\pi^2}{90}-
\frac{4\pi^2\ln 2}{15} + \frac{8135}{2304}\zeta_3\approx 1.45.
\eeq

The contribution to the Lamb shift induced by the three-loop polarization
insertion in the external photon is then given by

\beq
\Delta E_{pol}=-4\pi
Z\alpha|\psi_n(0)|^2p_3(\frac{\alpha}{\pi})^3\delta_{l0}.
\eeq

We easily obtain

\beq
\Delta E_{pol}=-\frac{\alpha^3(Z\alpha)^4}{\pi^3 n^3}\:m\:
\left(\frac{m_r}{m}\right)^3\cdot 4 p_3
\approx -1.45\frac{\alpha^3(Z\alpha)^4}{\pi^3 n^3}\:m\:
\left(\frac{m_r}{m}\right)^3
\eeq

or numerically

\beq
\Delta E(1S)_{pol}\approx -6.36 \mbox{ kHz},
\eeq
\[
\Delta E(2S)_{pol}\approx -0.79 \mbox{ kHz}.
\]
for the shifts in hydrogen.

\section{One More Low Energy Logarithm}

There is a very simple single logarithmic contribution of order
$\alpha^2(Z\alpha)^6$ connected with the irreducible two-loop vacuum
polarization.  The respective logarithmic term connected with the one-loop
vacuum polarization is well known \cite{lay,fy} and is equal to

\beq
\Delta E=-\frac{2}{15}\frac{\alpha(Z\alpha)^6}{\pi
n^3}\ln(Z\alpha)^{-2}(\frac{m_r}{m})^3m.
\eeq

This logarithm originates from the logarithmic correction to the
Schr\"odinger-Coulomb wave function which emerges when one takes
into account the one-photon exchange potential in the Dirac equation.
In other words, this is
exactly the logarithm which arises when one expands the Dirac-Coulomb wave
function in $Z\alpha$ and takes into account that we consider now
small distances of order $1/m$. In terms of the corrections to the
Schr\"odinger-Coulomb wave function we discuss the term

\beq
\delta\phi=-\frac{1}{2}(Z\alpha)^2\log Z\alpha \cdot\phi.
\eeq

Hence, the numerical factor in the contribution to the energy level
above is the product of this correction to the wave-function,
of the leading term in the low-frequency asymptotic expansion of the
one-loop polarization operator $-1/15$, and of the factor $4\pi(Z\alpha)$
which survives from the one photon exchange. One has also to take into
account a factor of two corresponding to the corrections to both wave functions
in the matrix element.

Now it is easy to obtain the respective result for the two-loop polarization.
The low-frequency asymptotic behavior of the two-loop polarization is described
by the factor $-41/162$ (see, e.g., \cite{sch}). Hence, we obtain (it is
necessary to also include an extra factor $\alpha/\pi$)

\beq
\Delta E_{log}=-\frac{2\cdot 41}{162}\frac{\alpha^2(Z\alpha)^6}{\pi^2
n^3}\ln(Z\alpha)^{-2}(\frac{m_r}{m})^3m
\eeq
\[
=-\frac{41}{81}\frac{\alpha^2(Z\alpha)^6}{\pi^2
n^3}\ln(Z\alpha)^{-2}(\frac{m_r}{m})^3m.
\]

or numerically

\beq
\Delta E(1S)_{log}\approx -0.50 \mbox{ kHz},
\eeq
\[
\Delta E(2S)_{log}\approx -6.26\cdot 10^{-2} \mbox{ kHz}.
\]

It is easy to see that correction of order $\alpha^3(Z\alpha)^4$
obtained above is small but nonnegligible as compared with current
experimental accuracy of  the $1S$ Lamb shift measurements (see, e.g.,
discussion in \cite{es}), while correction of order $\alpha^2(Z\alpha)^6$ is
too small to be of any phenomenological significance now.  In view of
the smallness of the above logarithmic correction, it is presumed that
any nonlogarithmic terms of this order would likewise be too small.

\acknowledgements

The research described in this publication was supported in part by the
National Science Foundation under grant \#NSF-PHY-9421408. Work of M.E.
was supported in part by grant \#R2E300 from the International Science
Foundation and was also supported by the Russian Foundation for Fundamental
Research under grant \#93-02-3853.

\end{document}